\documentclass{jltp}

\usepackage{graphicx}

\author{Ann Sophie C. Rittner and John D. Reppy}

\address{Laboratory of Atomic and Solid State Physics \\
and the Cornell Center for Materials Research,\\
Cornell University, Ithaca, New York 14853-2501}

\runninghead{A. S. C. Rittner and J. D. Reppy}{The annealing process
in solid $^4$He}

\title{Annealing in solid $^4$He}

\begin{document}

\maketitle

\begin{abstract}
We have used a torsional oscillator with square cross section and a
resonance frequency of 185 Hz to confirm the nonclassical rotational
inertia (NCRI) discovered by Kim and Chan\cite{1,2}. We have also
found a strong correlation between the NCRI signal and a high
dissipation Q$^{-1}$ of 4 $\times$ 10$^{-6}$ of the oscillation
above the transition temperature. Here, we present preliminary
results of the annealing process in $^{4}$He at a pressure of 26
bar. When holding the temperature constant above 1 K we have
observed a immediate rise in the period and a slow decay of the
dissipation. The equilibrium value of Q$^{-1}$ decreases with
increasing temperature.

\end{abstract}

\section{Introduction}
Eunseong Kim and Moses Chan (KC) of Penn State University first
observed \cite{2} nonclassical rotational inertia in solid $^4He$
(NCRI) below 250 mK. As previously reported \cite{wir}, we were able
to confirm NCRI in bulk solid. More importantly, we also observed
classical rotational inertia in the same cell by annealing the
sample and by raising the pressure above 32 bar.

All crystals in the NCRI state had a dissipation $Q^{-1}$ above
$4\times10^{-6}$ at 300 mK. The classical rotational inertia state
(CRI) was characterized by a relatively low dissipation of
$1\times10^{-6}$, comparable to the empty cell dissipation. We
believe that this low dissipation could indicate a state with less
defects than in the supersolid samples. This observation suggests
that crystal imperfections like grain boundaries or point defects
are probably essential for the existence of the supersolid state.

\section{Experimental Setup}
\begin{figure}
\setlength{\unitlength}{1.0in}
\begin{picture}(2.5,2.5)
\put(1.25,0) { \makebox(2.5, 2.5)[t] {
\includegraphics[width=0.7\textwidth]{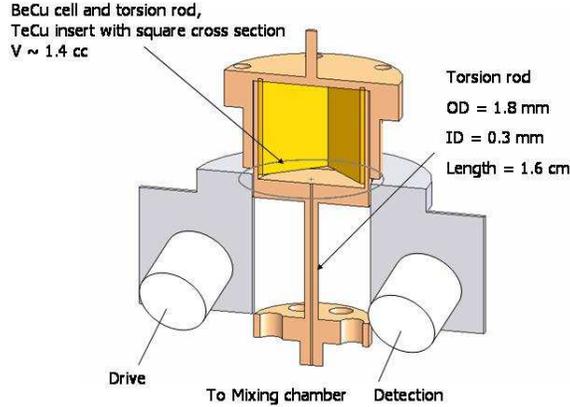}}
}
\end{picture}
\caption{\label{1oscillator}Torsional oscillator: The torsion cell's
motion is excited and detected capacitively. The AC voltage on the
detection electrodes serves as reference signal for a lock-in
amplifier to keep the oscillation in resonance. At 4 K, the
mechanical quality factor is 9 x 10$^5$, and the resonance frequency
is 185 Hz.}
\end{figure}

A schematic view of the oscillator is shown in Fig.\
\ref{1oscillator}. We used an oscillator with a volume of 1.4
cm$^{3}$, operated at 185 Hz. The scale for the dissipation $Q^{-1}$
is determined by a ring-down measurement of Q at the base
temperature of 20 mK. The empty cell value at 1 K is 1.0 $\times$
10$^{-6}$. A

The interior of the cell is a nearly cubic geometry which was
obtained by epoxying a Tellurium-Copper insert into the cylindrical
volume of the Beryllium-Copper oscillator. When cooling down, the
solid samples were grown with the blocked capillary method such that
the helium crystal was formed in the fill line first. The helium
used in the experiments had a nominal $^3$He concentration of 0.2 -
0.3 ppm. The pressure of the solid samples, discussed here, was 26
bar.

\section{Experimental observations}

After forming the solid sample, we cooled it down to 20 mK. Below
250 mK, we observed that the period gradually falls below the empty
cell value with the biggest change below 200 mK. Below 100 mK, the
period levels off. The total period drop is 80 ns, which compares to
the NCRI that KC as well as Motoshi Kondi and Keiya
Shirahama\cite{9} observe. Period and dissipation signals for a
typical cooldown at this pressure and velocity are shown
elsewhere\cite{wir}.

\begin{figure}
\centerline{\includegraphics[height=3in]{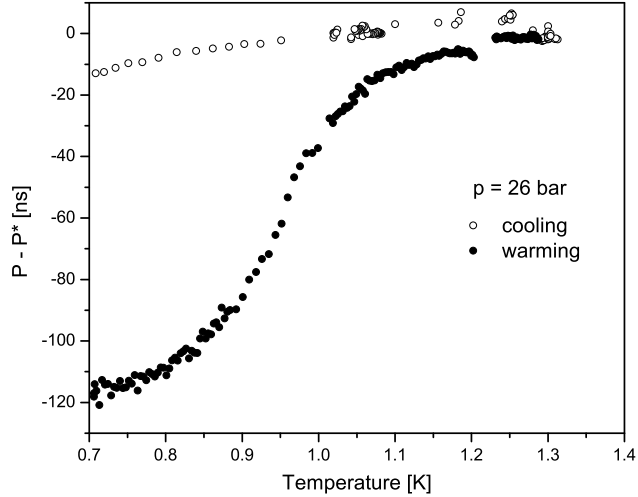}} \caption{The
resonant period while warming up (closed circles) and cooling down
(open circles) is shown as a function of temperature for a sample at
26 bar. The data was taken while warming up slowly after a cool-down
that showed NCRI with a period drop of 80 ns. The maximum wall
velocity is 80 $\mu$m/s.} \label{period}
\end{figure}

\begin{figure}
\centerline{\includegraphics[height=3in]{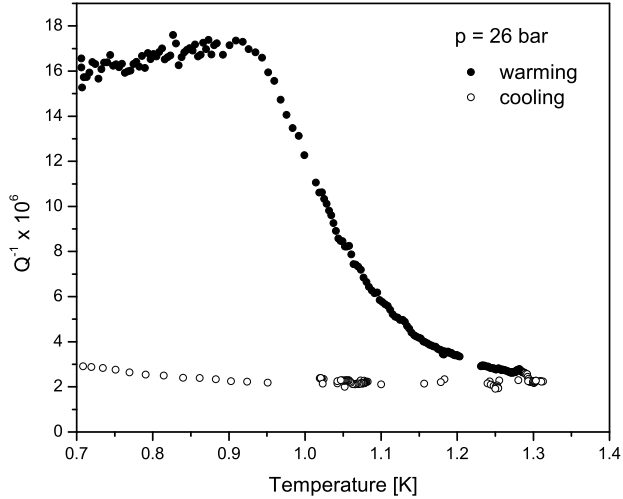}}
\caption{The dissipation while warming up (closed circles) and
cooling down (open circles) is shown as a function of temperature
for a sample at 26 bar. The data was taken while warming up after a
cool-down that showed NCRI and $Q^{-1}$[300 mK] of 16$\times$
$10^{-6}$. The maximum wall velocity is 80 $\mu$m/s.}
\label{dissipation}
\end{figure}

Afterwards, we warmed up very slowly such that the sample was held
between 1.0 K and 1.3 K for 16 hours. The period and dissipation
(both filled circles) $Q^{-1}$ as a function of temperature are
shown in Fig.\ \ref{period}, \ref{dissipation}. At temperatures
above 0.8 K, the period rises steeply by 100 ns until it levels off
at about 1.2 K. In the same temperature interval, the dissipation
decays from 16 $\times$ 10$^{-6}$ (at 0.7 K) to 2 $\times$ 10$^{-6}$
(at 1.3 K). This contrasts the behavior between 0.3 K and 0.8 K
where we have observed no sudden changes in either P or Q$^{-1}$.
Since the warm-up was very slow, we believe that the period and
dissipation were in thermal equilibrium.

After having held the temperature between 1.2 K and 1.3 K for nearly
12 hours, we lowered the temperature again below 1 K. The period and
dissipation (open circles) of this cool-down are also shown in Fig.
\ref{period}, \ref{dissipation}. When the temperature is lowered the
curves do not reverse. Instead, the period drop is reduced to 10 ns
and the dissipation exhibits only a small drop to 3 $\times$
10$^{-6}$ at 0.7 K. We cooled down through the supersolid transition
and observed a greatly decreased period drop of 7 ns at 170 mK. This
is the ''partially annealed" NCRI as reported earlier\cite{wir}.

\begin{figure}
\centerline{\includegraphics[height=2.85in]{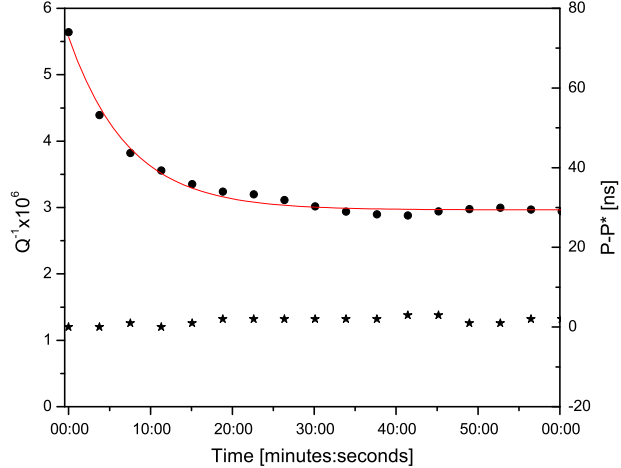}} \caption{The
dissipation data (circles) and resonant period (stars) are shown as
a function of time at 1.2 K. The data was taken after a cool-down
with a period drop of 50 ns at a pressure of 26 bar. The period is
shifted by P* ( = 5.428337 ms), the resonant period at 1.2 K. The
dissipation data is fit exponentially (red line) and yields a time
constant of 8 minutes.} \label{decay}
\end{figure}

Another sample that showed NCRI at low temperatures was warmed up
quickly to 1 K. Between 250 mK and 900 mK the dissipation was of the
order of 10 $\times$ 10$^{-6}$ without any sudden changes and the
period increased only gradually. We also observed a gradual increase
of the empty cell period when warming. After reaching 1 K, we warmed
up in steps, holding the temperature for about an hour each at 1 K,
1.2 K and 1.44 K. The melting temperature of $^4$He at 26 bar is 1.5
K. In Fig.\ \ref{decay}, we show an example of dissipation and
period versus time at a temperature of 1.2 K. When the temperature
is held constant, the period rises immediately (not shown in Fig.\
\ref{decay}) whereas the dissipation decays slowly to an equilibrium
value Q$^{-1}_0$. We fit the data to an exponential decay and obtain
a time constant of about 8 minutes. This time constant is similar
for the three temperatures. With increasing temperature, the period
increases which confirms the temperature dependence observed in the
sweep (Fig. \ref{period}). The dissipation drops from 8.6 $\times$
10$^{-6}$ (1 K) to 2.9 $\times$ 10$^{-6}$ (1.2 K) and finally 2.5
$\times$ 10$^{-6}$ (1.44 K). The period rises from 5.428310 ms (1 K)
by 27 ns (1.2 K) and 45 ns (1.44 K) respectively. The equilibrium
values for period and dissipation match the ones found in Fig.
\ref{period}, \ref{dissipation}.

\section{Discussion}
We find that the NCRI is associated with a state of high
dissipation, Q$^{-1}$, for the $^4He$ solid which may be caused by
defects in the crystal. The absence of supersolid behavior in the
annealed samples is supported by the recent theoretical findings
that an ideal hcp $^{4}$He crystal does not support ODLRO or BEC
\cite{7,8}. When we warm up and anneal the sample the dissipation
decays to almost the empty cell value of 1.0 $\times$ 10$^{-6}$. One
may explain this drop with defects or dislocations leaving the
crystal.

KC and Shirahama have also undertaken annealing experiments
\cite{9,10}, but have not been able to eliminate their NCRI signals.
We note that there are differences in geometry, pressure and
frequency between the experiments. At the moment, we do not know how
the experimental differences contribute to crystal quality and to
successful annealing.

We also find that the dissipation does not decay to its lowest value
of about 1 $\times$ 10$^{-6}$ at temperatures at and below 1.44 K.
Instead, the equilibrium value for the dissipation is temperature
dependent such that higher temperatures are needed in order to
completely remove the NCRI signal.

\section*{ACKNOWLEDGMENTS}
The work reported here has been supported by the National Science
Foundation under Grant DMR-060584 and through the Cornell Center for
Materials Research under Grant DMR-0520404.

\end{document}